\documentstyle[12pt]{article}  

\def\cal{\mathcal}
\newcommand{\comma}{\;\;\; ,}
\newcommand{\period}{\;\;\; .}

\newcommand{\eq}{\; = \;}
\newcommand{\sep}{\;\; , \;\;}
\newcommand{\be}{\begin{equation}}
\newcommand{\bd}{\begin{displaymath}}
\newcommand{\ed}{\end{displaymath}}
\newcommand{\ba}{\begin{eqnarray}}
\newcommand{\ea}{\end{eqnarray}}
\newcommand{\Wb}{\overline{W}}
\newcommand{\Yb}{\overline{Y}}
\newcommand{\Xb}{\overline{X}}
\newcommand{\Pb}{\overline{P}}
\newcommand{\fb}{\overline{f}}
\newcommand{\Kb}{\overline{K}}
\newcommand{\Thb}{\overline{\Theta}}
\newcommand{\omegb}{\overline{\omega}}
\newcommand{\journalnumber}{\bf}

\begin{document}
\pagenumbering{arabic}
   \title{~~~~~~A rapidity-independent  parameter \hfil\break in the  
   star- triangle relation }
\maketitle

\begin{center}
\author{ R.J. Baxter\\
}
\end{center}


{\renewcommand{\thefootnote}{\fnsymbol{footnote}}

\footnotetext{\kern-19pt{\bf AMS Subject classification:} Primary
82B20; secondary
82B23. }

\footnotetext{\kern-19pt{\bf  Keywords and phrases:} statistical mechanics,
   lattice models, star-triangle relation, exactly solvable models,
   Kashiwara-Miwa model,  chiral Potts model }


\begin{abstract}
The normalization factor in the star-triangle relation can be evaluated in a simple form by
taking determinants. If we combine this with the rotation
symmetries, then we can show that a certain simple quantity $I$ has to be
independent of the rapidities. In this sense it is an invariant. We evaluate it
for several particular models and find it is one for self-dual models, and is
related to the modulus $k$ (or $k'$) for the Ising, Kashiwara-Miwa and chiral
Potts models.
\end{abstract}
~~~
\\
{\bf {This paper is intended as a contribution to the volume dedicated
to the sixtieth birthday of
Barry M. McCoy, which event was marked in Okayama and Kyoto in February 2001.}}
\\
~~~\\
23 August 2001

\section{Introduction}

I have known Barry McCoy since 1972, when I spent six months at Stony Brook.
His enthusiasm for theoretical physics is infectious, and is coupled with a
great tenacity in tackling long and extremely hard calculations. I am still 
impressed by the evaluation he did then, with Johnson and Krinsky,
of the correlation length of the eight-vertex model.\cite{Johnson} In this calculation it is not
enough merely to find the next-largest eigenvalue of the transfer
matrix. Instead one has to determine a whole band of eigenvalues, then perform the sum over this
band. In the thermodynamic limit this sum becomes an integration, and one evaluates it by a 
saddle-point method. A tour de force!

Much of the work in solvable models stems from the Yang-Baxter relation, in particular 
its star-triangle form. Here I shall show how one can quite easily use this
relation to  show that a certain expression $I$ is invariant, by which I mean that it
is independent of the  rapidities. It is therefore a temperature-like variable, being a
function only of the modulus and crossing parameter. For self-dual $Z_N$ invariant models $I$ is
one. 

We evaluate $I$ for the Ising, self-dual Potts, self-dual Ashkin-Teller,\\ Kashiwara-Miwa,
chiral Potts and Fateev-Zamolodchikov models. In all cases it is either one, or is related to
the modulus $k$ (or $k'$).

\pagestyle{myheadings}
\markboth{R.J. Baxter}{The Star-Triangle Relation}

\section{Star-triangle relation} 

 Consider a model on the square lattice where spins take values $0,...,N-1$.
Two adjacent spins $a$ and $b$ interact with a Boltzmann weight function
$W_{pq}(a,b)$ if $b$ is horizontally to the right of $a$, and $\Wb_{pq} (a,b)$ if 
$b$ is vertically above $a$, as in the figure. As usual, $p$ and $q$ are
rapidity variables, associated with the broken lines.


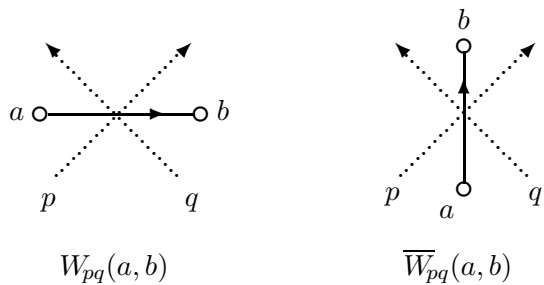
\typeout{  }
\typeout{     NB:   There follow four error messages about ovals, }
\typeout{     circles and lines caused by the small vectors }
\typeout{     in these LATEX PICTURE MACROS  , but  }
\typeout{					they are harmless for Rodney Baxter's latex systems  }
\typeout{					and I do get the pictures I want.     }
\typeout{				If they really annoy you, see the Latex code for the }
\typeout{				     LATEX PICTURE MACROS PROBLEM:    }

\setlength{\unitlength}{.15in}
\thicklines
\def\punit#1{\hspace{#1\unitlength}}
\def\pvunit#1{\vspace{#1\unitlength}}
\def\Wpqfig#1#2#3#4{\rule[-2.8\unitlength]{0in}{5.6\unitlength}
\begin{picture}(4,8)(0,-4.0)
\put(-6.5,0){\line(1,0){5.1}}
\put(-3.4,0){\vector(1,0){1.0}}
\put(-6.5,-3.2){\makebox(0,0)[b]{\small \mbox{$#1$}}}
\put(-1.5,-3.2){\makebox(0,0)[b]{\small \mbox{$#2$}}}
\put(-0.4,-0.2){\makebox(0,0)[b]{\small \mbox{$#3$}}}
\put(-7.6,-0.2){\makebox(0,0)[b]{\small \mbox{$#4$}}}
\put(-4.2,-5.6){\makebox(0,0)[b]{\small \mbox{$W\! _{\! pq} (a, b)$}}}
\multiput(-6.4,-2.2)(0.2,0.2){22}{\makebox(0,0)[b]{ \mbox{${.}$}}}
\multiput(-6.4,2.0)(0.2,-0.2){22}{\makebox(0,0)[b]{ \mbox{${.}$}}}
\put(-6.8,0){\circle{.5}}
\put(-1.2,0){\circle{.5}}
\put(8.0,-2.4){\line(0,1){4.6}}
\put(8.0,0.3){\vector(0,1){1.0}}
\put(5.5,-3.0){\makebox(0,0)[b]{\small \mbox{$#1$}}}
\put(10.5,-3.0){\makebox(0,0)[b]{\small \mbox{$#2$}}}
\put(8.0,3.0){\makebox(0,0)[b]{\small \mbox{$#3$}}}
\put(7.4,-3.6){\makebox(0,0)[b]{\small \mbox{$#4$}}}
\put(7.8,-5.6){\makebox(0,0)[b]{\small \mbox{$\overline{W}\! _{\! pq} (a, b)$}}}
\multiput(5.8,-2.2)(0.2,0.2){22}{\makebox(0,0)[b]{\mbox{${.}$}}}
\multiput(5.8,2.2)(0.2,-0.2){23}{\makebox(0,0)[b]{\mbox{${.}$}}}
\put(8.0,-2.6){\circle{.5}}
\put(8.0,2.4){\circle{.5}}




\put(-1.8,2.2){\vector(1,1){0.3}}
\put(-6.3,2.2){\vector(-1,1){0.3}}
\put(10.2,2.2){\vector(1,1){0.3}}
\put(5.9,2.2){\vector(-1,1){0.3}}

\end{picture}}

\bd  \Wpqfig{p}{q}{b}{a} \ed

To be as general as possible, we should also allow a rapidity-independent field weight
$S(a)$ for each spin $a$ on the lattice\footnote{It may be possible
to absorb $S(a)$ into $W_{pq}(a,b)$ and/or 
 $\Wb_{pq}(a,b)$, but to the author it seems  clearer not to do so.}.
 Then the star-triangle relation is:
\ba  \label{startri}
\sum_{d} S(d) \, \overline{W}_{qr}(a,d)  W_{pr}(c,d) 
\overline{W}_{pq}(d,b) & \! \! =  \! \! &
{\cal R}_{pqr}\,  W_{pq}(c,a)  \overline{W}_{pr}(a,b) 
W_{qr}(c,b) \nonumber \\ 
&& \\
\sum_{d} S(d) \, \overline{W}_{qr}(d,a)
W_{pr}(d,c)  \overline{W}_{pq}(b,d) &  \! \! =  \! \!  &
{\cal R}_{pqr}\, W_{pq}(a,c) \overline{W}_{pr}(b,a)
W_{qr}(b,c)  \nonumber \ea
	where  $ {\cal R}_{pqr}$ is some factor independent of the
spins $a, b, c$.

It is helpful to consider a honeycomb lattice with edge weights $\overline{W}_{qr}$,
$W_{pr}$, $\overline{W}_{pq}$, and a triangular lattice with weights $ W_{qr}$,
$\overline{W}_{pr}$, $W_{pq}$, both with site weights $S$. Then the two relations 
(\ref{startri}) are the conditions for them to be related, working from either down-pointing or
up-pointing stars or triangles.

For most of the solved two-dimensional models the weight functions $W_{pq}(a,b)$  and $\Wb_{pq}
(a,b)$  are symmetric, {\it i.e.} unchanged by interchanging the spin arguments $a$ and $b$.
Clearly the two relations (\ref{startri}) are then equivalent. The only asymmetric model
satisfying the star-triangle relation known to the author  is the chiral Potts
model \footnote{There are other solvable planar models, derived from the
three-dimensional  Zamolodchikov model, which are asymmetric, but these satisfy the
``star-star'' relation, rather than the star-triangle.\cite{Bax97}}, but for this model
$S(a) = 1$ and $W_{pq}(a,b)$, $\Wb_{pq}(a,b)$ depend on
$a$, $b$ only via their difference $a-b$, modulo $N$. It follows that  one can convert one of the
relations (\ref{startri}) into the other by negating (modulo $N$) all spins.

\typeout{    *******  See above re four error messages about ovals, ****** }
\typeout{    *******       circles and lines                        ****** }
\typeout{  }

Thus the two relations (\ref{startri})  seem to be  equivalent for all known models. We
shall not in fact use this, but we shall use a property that is certainly true of all the
models considered here, namely that there exist spin-independent quantities $P_{pq}$,
$\Pb_{pq}$ such that
\ba \label{Pdef}
\prod_{b} W_{pq}(a,b)  \eq \prod_{b} W_{pq}(b,a) & = & P_{pq} \comma \nonumber \\
\prod_{b} \Wb_{pq}(a,b)  \eq \prod_{b} \Wb_{pq}(b,a) & = & \Pb_{pq} \comma \ea
for all values of the spin $a$. 
let $Y_{pq}$ be the $N$ by $N$ matrix with entry
$W_{pq}(a,b)$ in position $(a, b)$, and $\Yb_{pq}$ the matrix with
corresponding entry
$\Wb_{pq}(a,b)$. Then the properties (\ref{Pdef}) state that the row and column products of
$Y$ are equal to one another and are the same for all rows and columns. Similarly for $\Yb$.
It is probably not essential to assume these properties: it may be that they are implied by
(\ref{startri}), to within a gauge transformation. However, they are in fact satisfied for
our specific models, and they do greatly simplify the following argument.

In addition to the matrices $Y_{pq}$, $\Yb_{pq}$, we shall also need the diagonal matrices
$S$, $X_{pq|c}$, $X'_{pq|c}$,$\Xb_{pq|c}$, $\Xb'_{pq|c}$ with entries
$S(a) \delta(a,b)$, $W_{pq}(c,a) \delta(a,b)$, $W_{pq}(a,c) \delta(a,b)$, 
$\Wb_{pq}(c,a) \delta(a,b)$, $\Wb_{pq}(a,c) \delta(a,b)$ in position $(a, b)$, respectively.
Note that  $X_{pq|c}$, $X'_{pq|c}$, $\Xb_{pq|c}$, $\Xb'_{pq|c}$ depend on the spin $c$.

We also define
\be s \eq [S(0) \cdots S(N-1)]^{1/N} \period \end{equation}

\section{The factor  $ {\cal R}_{pqr}$ }

For the chiral Potts model, Baxter, Perk and Au-Yang \cite{BPAY88} conjectured 
the form of the
factor $ {\cal R}_{pqr}$. This conjecture was verified by  
Matveev and Smirnov \cite{MatSmirn90}: their argument generalizes at once 
to any model satisfying
(\ref{startri}), as we shall now show.

Regard the spin $c$ as fixed and think of
each side of (\ref{startri}) as the element $(a,b)$ of some matrix. Then the  relations
take the matrix form 
\ba \label{mateqns}
\Yb_{qr} \, S X_{pr|c} \, \Yb_{pq} & \eq & {\cal R}_{pqr} \, X_{pq|c} \, \Yb_{pr} \, X_{qr|c}
\nonumber
\\  \Yb^T_{qr} \, S X'_{pr|c} \, \Yb^T_{pq} & \eq & {\cal R}_{pqr} \, X'_{pq|c} \, 
\Yb^T_{pr} \, X'_{qr|c} \comma \ea
$\Yb^T$ being the transpose of $\Yb$. Since $c$ takes $N$ values, (\ref{mateqns}) consists 
of $2 N$ separate matrix equations.

One can now obtain ${\cal R}_{pqr}$ by taking determinants. Because of (\ref{Pdef}) we
obtain not $2 N$ results, but only one, namely
\be \label{result1}
{\cal R}_{pqr} =  f_{qr} f_{pq}/f_{pr}  \end{equation}
(to within an undetermined factor of an $N$th root of unity). 

Here $f_{pq}$ is defined by
\be \label{deffpq}
 f^N_{pq} \eq P_{pq}^{-1} \, {\rm det} \, (S   \Yb_{pq} )   \period   \end{equation}
When $S(a) = 1$ we regain the formulae of \cite{BPAY88,MatSmirn90}.

\section{The invariant  $I$}

 Define also
$\fb_{pq}$ by
\be \label{deffbpq}
 \fb^N_{pq} \eq \Pb_{pq}^{-1} \, {\rm det} \, (S  Y_{pq})  \period \end{equation} 

We obtained (\ref{result1}) by holding {$c$} fixed and regarding each side of the
star-triangle relations as  elements $(a,b)$ of a matrix product. Suppose instead we hold
{$a$} fixed and think of them as elements $(c,b)$. Then the resulting matrix relation is
\ba
Y_{pr} \, S \Xb_{qr|a} \, \Yb_{pq} & \eq & {\cal R}_{pqr} \, X'_{pq|a} \, Y_{qr} \,
\Xb_{pr|a}
\nonumber
\\  Y^T_{pr} \, S \Xb'_{qr|a} \, \Yb^T_{pq} & \eq & {\cal R}_{pqr} \, X_{pq|a} \, 
Y^T_{qr} \, \Xb'_{pr|a} \period \ea
Taking determinants, these give (for all $a$ and to within an $N$th root of unity)
\be \label{result2}
{\cal R}_{pqr} \eq  f_{pq} \, \fb_{pr}/\fb_{qr} \period \end{equation}

Finally, holding {$b$}  fixed and regarding the relations as elements $(c,a)$ 
of a matrix, we get
\ba
Y_{pr} \, S \Xb '_{pq|b} \, \Yb^T_{qr} & \eq & {\cal R}_{pqr} \, X'_{qr|b} \, Y_{pq} \, 
\Xb'_{pr|b} \nonumber
\\  Y^T_{pr} \, S \Xb_{pq|b} \, \Yb_{qr} & \eq & {\cal R}_{pqr} \, X_{qr|b} \, 
Y^T_{pq} \, \Xb_{pr|b} \comma \ea
\be \label{result3}
{\cal R}_{pqr} \eq  f_{qr} \, \fb_{pr}/\fb_{pq} \period \end{equation}

The relations (\ref{result1}), (\ref{result2}), (\ref{result3}) are 
mutually consistent if and only if
\be
f_{pq} \, \fb_{pq} \eq f_{pr} \, \fb_{pr} \eq f_{qr} \, \fb_{qr} \period \end{equation}
The only way this can happen is for the quantity
\be \label{defI}
I \eq N^{-1} \, f_{pq} \, \fb_{pq}  \end{equation}
to be {\em independent of $p$ and $q$ }. (We have introduced the factor $1/N$ for later
convenience.)

Thus it follows, for any model satisfying the star-triangle relations and having 
the properties (\ref{Pdef}), that
 $I$ is an invariant, independent of the rapidities $p$ and $q$. It is independent of the
normalization of $W_{pq}(a,b)$,  $\Wb_{pq}(a,b)$, i.e. it is unchanged by multiplying
them by factors independent of the spins $a, b$.

\section{Inversion and rotation relations}

All of the solvable models we shall discuss can be normalized  so 
as to have the properties
\be
W_{pq} (a,b) W_{qp} (a,b) \eq 1 \sep \Wb_{pp} (a,b) \eq \delta(a,b) /S(a) \period \end{equation}
In particular, $W_{pp}(a,b) = P_{pp} = f_{pp} = 1$. Setting $r = p$ in (\ref{mateqns}),
then interchanging
$p$ with $q$ and using (\ref{result1}), it follows that
\be \label{invsn}
\Yb_{pq} \, S \, \Yb_{qp} \eq (f_{pq} f_{qp} ) \, S^{-1}  \period \end{equation}

Further, there always exists mapping $R: p \rightarrow Rp$ such that
\be \label{rotn}
\Wb _{pq} (a,b) \eq W_{q,Rp} (a,b) \sep W_{pq} (a,b) \eq \Wb_{q,Rp} (b,a) \period \end{equation}
This means that replacing $p, q$ by $q, Rp$ is equivalent to rotating the lattice
through $90^{\circ}$. For all the models except the chiral Potts model, the mapping is
very simple: $Rp = p + \lambda$, where $\lambda$ is a fixed ``crossing
parameter''. These relations (\ref{rotn}) imply that
\be \label{frotn}
\fb _{pq} \eq f_{q,Rp} \period \end{equation}

If $\kappa_{pq}$ is the partition function per site in the large-lattice limit, then
it follows that
\be
\kappa_{pq} \eq \kappa_{q,Rp} \eq f_{pq} f_{qp}/\kappa_{qp} \period \end{equation}
Together with an appropriate analyticity assumption about $\kappa_{pq}$
in a domain containing the inversion points $(p,p)$ and $(p,Rp)$, these
relations can be used (at least for the models other than the chiral Potts model)
to obtain  $\kappa_{pq}$ \cite{Stroganov, Bax82}.

\section{$Z_N$-invariant models: duality}

Three of the models we shall discuss, namely the Ising, self-dual Potts and
chiral Potts, are $Z_N$ invariant. That is, $S(a) = 1$ and 
\be \label{zinvww}
W_{pq} (a,b) \eq W_{pq} (a - b) \sep \Wb_{pq} (a,b) \eq \Wb_{pq} (a - b) \comma \end{equation}
the spin-difference functions $ W_{pq} (i)$, $\Wb_{pq} (i)$ being periodic in $i$ of
period $N$. The matrices $Y_{pq}$, $\Yb_{pq}$ are therefore cyclic (Toeplitz).

In this case there is a simple duality property \cite[\S 6.2]{book}, \cite[\S
5]{Bax89}.  To within boundary conditions, the partition function of the
square lattice model is unchanged by replacing
$ W_{pq} (j)$, $\Wb_{pq} (j)$ by the Fourier transforms
\ba
W^{(d)}_{pq} (j) & = & N^{-1/2} \, \sum_{n= 0}^{N-1} \omega^{j n} \Wb_{pq} (n) \comma
\nonumber \\
\Wb^{(d)}_{pq} (j) & = & N^{-1/2} \, \sum_{n= 0}^{N-1} \omega^{-j n} W_{pq} (n) \comma \ea
where $\omega = \exp ( 2 \pi i /N)$ is the primitive $N$th root of unity.

Using the fact that the determinant of a matrix is the product of its eigenvalues,
which for the Toeplitz matrices $\Yb_{pq}$, $Y_{pq}$ are the above sums, it follows that
\be
I^N \eq P^{(d)}_{pq} \, \Pb^{(d)}_{pq} / \left( P_{pq} \Pb_{pq} \right) \comma \end{equation}
$P^{(d)}_{pq}$, $\Pb^{(d)}_{pq}$ being defined by (\ref{Pdef}) with $W, \Wb$ replaced by 
the duals  $W^{(d)}$, $\Wb^{(d)}$.

Making appropriate choices of $N$th roots, it follows that $I$ is inverted by such a
duality transformation, and at the self-dual point
\be \label{selfdual}
I = 1 \period \end{equation}

\section{Particular Models}

\subsection*{Ising model}

For the Ising model with horizontal and  vertical interaction coefficients
$J, \overline{J}$, $N=2$ and
\be
Y_{pq} \eq  \left(\begin{array}{cc}
					               e^K & e^{-K} \\  e^{-K} &  e^K \end{array} \right) \sep
\Yb_{pq} \eq  \left(\begin{array}{cc}
					               e^{\Kb} & e^{-\Kb} \\  e^{-\Kb} &  e^{\Kb} 
\end{array} \right) \comma
\end{equation}
where $K = J/k_B T$, $\Kb = \overline{J}/k_B T $. Hence
\be \label{IsingI} 
I^2 = \sinh 2K \, \sinh 2 \Kb \period \end{equation}
This is indeed the basic invariant of the  Ising model - being the modulus
$1/k$ or $k$ of its elliptic function parametrization
 \cite[eqns. (2.1a) and (2.1b)]{Onsager}, \cite[eqns. (3.67) and (3.71)]{McCoyWu73}.
For
$I^2 >1$ the model is ferromagnetically ordered, for $I^2 < 1 $ it is disordered, and
when
$I^2 = 1$ it is critical. The Boltzmann weights $e^{-2K}, e^{-2\Kb}$ are functions of
the rapidity variables $p, q$, in fact elliptic functions of $p - q$. We shall not
pursue this further as the Ising model is the $N = 2$ case of the Kashiwara-Miwa and
chiral Potts models discussed below.

\subsection*{Self-dual Potts model}

 The ordinary Potts model
\cite{Potts52} is an $N$ (or $q$) - state model with weights $S(a) = 1$,
\be 
 W(a,b) \eq e^{-K}+ (1 - e^{-K}) \delta_{ab}
 \sep \Wb(a,b) \eq e^{-\Kb}+ (1 - e^{-\Kb})  \delta_{ab} \period \end{equation}
It is solvable at its self-dual point, i.e. when
\be \label{selfdualP}
(e^K - 1) \, (e^{\Kb} - 1) \eq  N \comma \end{equation}
when it is equivalent to the six-vertex model \cite[\S 12.3]{book}.
At this point the rapidity parametrization is
\be
e^K \eq \frac{\sin (\mu+q-p)}{\sin (\mu-q+p )} \sep
e^{\Kb} \eq \frac{\sin (2 \mu-q+p)}{\sin (q-p )}  \comma \end{equation}
where $\mu$, the crossing parameter, is defined by
\be
N^{1/2} \eq 2 \cos \mu   \period \end{equation}
For $N < 4$, $\mu,p, q $ are real. For $N > 4$, they are pure imaginary.
As $N \rightarrow 4$, they all become small but enter the equations only via their
ratios.

It follows that
\ba f_{pq} \eq e^K (1- e^{-\Kb}) & = & 2 \cos \mu \, \sin (\mu+q-p ) /
\sin (2 \mu -q+p) \nonumber \\
\fb_{pq} \eq e^{\Kb} (1- e^{-K}) & = & 2 \cos \mu \, \sin (2 \mu-q+p ) /
\sin (\mu +q-p ) \comma \ea
and hence
\be 
I \eq 1\comma \end{equation}
in agreement with (\ref{selfdual}).

\subsection*{Ashkin-Teller model}
The Ashkin-Teller model \cite{AshTell43} is usually formulated in terms of a pair of
two-valued spins at each site, interacting with neigbouring pairs via Ising and
four-spin interactions \cite[\S 12.9]{book}. This means that it is a four-state
model, with 
$S(a) = 1$ and interaction matrices
\be \label{wwKM}
Y_{pq} \eq \left(  \begin{array}{cccc}
\omega_0 & \omega_1 & \omega_2 & \omega_3 \\
\omega_1 & \omega_0 & \omega_3 & \omega_2 \\
\omega_2 & \omega_3 & \omega_0 & \omega_1 \\
\omega_3 & \omega_2 & \omega_1 & \omega_0 \\ \end{array} \right) \sep
\Yb_{pq} \eq \left(  \begin{array}{cccc}
\omegb_0 & \omegb_1 & \omegb_2 & \omegb_3 \\
\omegb_1 & \omegb_0 & \omegb_3 & \omegb_2 \\
\omegb_2 & \omegb_3 & \omegb_0 & \omegb_1 \\
\omegb_3 & \omegb_2 & \omegb_1 & \omegb_0 \\ \end{array} \right) \period \end{equation}
These matrices are not Toeplitz, but they are cyclic in two-by-two blocks. The 
restrictions (\ref{Pdef}) are obviously satisfied. The weights $\omega_0, \ldots
, \omegb_3$ can be expressed in terms of interaction coeficients, but we 
shall just take
them to be given parameters.

By performing a duality transformation on one of the two sets of Ising 
spins, Wegner \cite{Wegner72}, \cite[\S 12.9]{book} showed that this model can 
be converted to a staggered square-lattice eight-vertex model, with weights
\be (a,b,c,d) \eq (\omega_0 + \omega_1,
 \omega_2 - \omega_3, 
 \omega_2 + \omega_3, \omega_0 - \omega_1)/\sqrt{2} \end{equation}
on one sub-lattice, and weights
\be( \overline{a},\overline{b},\overline{c},\overline{d})
 \eq (\omegb_0 +\omegb_1, \omegb_2 - \omegb_3, 
 \omegb_0 - \omegb_1, \omegb_2 + \omegb_3)/\sqrt{2} \end{equation}
on the other.

This staggered model has not been solved and we know of no 
star-triangle relation relevant to it. However, it can be solved when the
two sets of eight-vertex weights are proportional to one another, as it is then 
the regular model: if
\be \label{sdAT}
(\overline{a}, \overline{b}, \overline{c}, \overline{d} ) = \xi (a, b, c,
d) \comma \end{equation} 
then $a,b,c,d$ can be parametrized as elliptic functions of $p - q$ so as to satisfy
(\ref{startri}). However, we do not need this parametrization to note from 
(\ref{wwKM}) - (\ref{sdAT}) that
\be
{\rm det} \, \Yb_{pq} \eq 16 \, \xi^4  \, P_{pq} \sep 
{\rm det} \, Y_{pq} \eq 16 \, \xi^{-4 } \,  \Pb_{pq} \comma \end{equation}
and hence ( to within choices of $N$th roots) $ f_{pq} = 2 \xi$, $\fb_{pq}
= 2/\xi$, and 
\be I \eq 1    \period \end{equation}

Thus $I = 1$, as for the self-dual $Z_N$-invariant models. In fact,
this solvable case of the Ashkin-Teller model {\em is} self-dual \cite{Wegner72,
WuLin74}, but it is {\em not} critical. We know that the associated regular
eight-vertex model can be naturally parametrized in terms of elliptic functions
of some modulus $k$ \cite{Baxter71,Baxter72}. Only when this $k$ is $\pm 1$ , $0$ or
$\infty$ can the model be critical.

The explanation of this apparent contradiction is  that the general Ashkin-Teller 
model has two
critical temperatures, on either side of the self-dual temperature defined by
(\ref{sdAT}), which map into one another by duality \cite{Wegner72, WuLin74}.
 (The duality relation is easily obtained from the above
remarks: two Ashkin-Teller models are dual to one another if they transform into staggered
eight-vertex models differing only in interchanging the two sub-lattices. This means that
the $\omega$-weights of one are proportional to the eigenvalues of the $\Yb , Y$
matrices of the other.)

Indeed, when the four-spin interaction vanishes the Ashkin Teller-model becomes two
independent Ising models. The solvability condition (\ref{sdAT}) then implies that these
two models are dual to one another, and $I$ is the product of the individual $I$s 
given by (\ref{IsingI}). This product is necessarily one.
 
\subsection*{Kashiwara-Miwa model}

  In 1986 Kashiwara and Miwa \cite{KashMiwa} presented an
$N-$state generalization of the Fateev-Zamolodchikov model that
breaks the $Z_N -$symmetry, but retains the reflection symmetry
and the rapidity difference property. This model has also been
studied by Hasegawa and Yamada \cite{HasYam}, and by Gaudin
 \cite{Gaudin}. Let  $K, K'$ be the complete elliptic integrals of
the first kind of moduli $k, k'$, and let  $\tilde{q} = \exp(-\pi 
K' /K) $ be the corresponding ``nome''. Then the  elliptic theta functions  $H(u)$, 
$\Theta (u)$ of argument $u$
and modulus $k$, as defined in section 8.181 of \cite{GR}, are 
\bd H(u) \eq   2 \tilde{q}^{1/4} \sin (\pi u /2K) \, \prod_{n = 1}
^{\infty} ( 1 - 2 \tilde{q}^{2n} \cos (\pi u /K)\, + 
\tilde{q}^{4n} )( 1 - \tilde{q}^{2n})  \comma \ed
\be \Theta (u) \eq  \prod_{n = 1}
^{\infty} ( 1 - 2 \tilde{q}^{2n - 1} \cos (\pi u /K) \, + 
\tilde{q}^{4n - 2} )( 1 - \tilde{q}^{2n})  \period \end{equation}

We also use the function $H_2 (u) = H(u) \Theta(u)$. One can write $K,K',
k, k'$ in terms of infinite products involving $\tilde{q}$, in particular
\be
\frac{2 k' K} {\pi} \eq \prod_{m=1}^{\infty} \left( 
\frac{1-\tilde{q}^m}{1+\tilde{q}^m}   \right)^2 \period \end{equation}

Let $\zeta$ be some arbitrary integer and define two functions
\be
r_v ( n)  \eq \prod_{j = 1} ^{n} \frac{H[K( 2  j
- 2  + v )/N]}{H[K(2  j - v  )/ N]}
\sep 
t_v ( n)  \eq \prod_{j = 1} ^{n} \frac{\Theta[K( 2  j
- 2  + v )/N]}{\Theta[K (2  j - v )/ N]}
 \comma \end{equation}
where $v$ is real and the argument $n$ is an integer. These functions are periodic:
$r_v ( n)  = r_v ( n + N) $, $t_v ( n)  = t_v ( n + N) $, 

Then the weights of the
Kashiwara-Miwa model are (for integers $a, b$ )
\be
S(a) = \Theta [2 K (2 a+\zeta)/N ]/ \Theta (0) \comma \end{equation}
\be W_{pq} (a,b)  \eq r_{1 - q + p } (  a-b) \; t_{1  - q + p } 
( a + b + \zeta)  \comma
\end{equation}
 \bd \Wb_{pq} (a,b)  \eq r_{q - p} ( a-b) \;
 t_{q - p } ( a + b + \zeta)  \period \ed 
Note that for this model,
unlike all the others we consider, the site weight
function $S(a)$ is {\em not} unity. Curiously, $  W_{pq} (a,b) $ and 
$\Wb_{pq} (a,b) $ are products of functions of the spin difference $a-b$ 
and of the spin sum $a+b$. Because $r_v (n) = r_v (-n)$ is an even function of $n$,
$  W_{pq} (a,b) $ and 
$\Wb_{pq} (a,b) $ are each unchanged by interchanging the spins $a, b$. The model is
therefore reflection-symmetric, i.e. non-chiral.

Define also
\be
g \eq \prod_{m=1}^{\infty} \frac{ 1 - \tilde{q}^{N m}}{1 + \tilde{q}^{N m}}
\; \frac{ 1 + \tilde{q}^{m }}{1 - \tilde{q}^{m }} \period \end{equation}
Then, based on numerical calculations, the known $N=2$ Ising case and the
Fateev-Zamolodchikov result (\ref{FZfpq}) below,  I conjecture that
\be \label{KMfpq}
f_{pq} \eq N^{1/2} \, g \, \prod_{j=1}^{N'} 
\frac{H_2 [K(q-p+2j-1)/N]}{H_2 [K(2j+p-q)/N]}   \comma     \end{equation} 
where $N' = [N/2]$ is the integer part of $N/2$. (The constant $g$ is determined by
the requirement that $f_{pp} = 1$.)

Since $\fb _{pq}$ is obtained from $f_{pq}$ by replacing $q-p$ by $1+p-q$, it then
follows that
\be \label{IKM}
I \eq g^2 \eq k_N' \, K_N/(k' K)  \comma \end{equation}
where $k_N', K_N$ are the 
$k', K$ corresponding to the elliptic nome $\tilde{q}^N$. This 
explicit expression for $I$ is indeed independent of the rapidities $p$, $q$.
Note also that $f_{pq}$ is a single-valued meromorphic function of $q-p$, despite the
$N$th root implied by the definition (\ref{deffpq}): this is consistent with the
observation \cite{MatSmirn90} that $f_{pq}$ is defined by (\ref{result1}) and 
(\ref{startri}) as a single-valued function of the Boltzmann weights (with 
{\em no} $N$th root), to within factors that are functions of either $p$ or $q$ only.

The factor $f_{pq} f_{qp}$ in (\ref{invsn}) also simplifies. Define the elliptic function
\be
G(u, \tilde{q} ) \eq \frac{H_2 (2K u/\pi)}{H_2 (K+2Ku/\pi)} \eq
\tan u \, \prod_{m=1}^{\infty} \frac{1-2 \tilde{q}^{\, m} \cos 2 u + \tilde{q}^{\, 2 m} }
{1+2 \tilde{q}^{\, m} \cos 2 u + \tilde{q}^{\, 2 m} } \period \end{equation}
Then
\be
f_{pq} f_{qp} \eq N \, g^2 \, G(w, \tilde{q} )/G(N w , \tilde{q}^N ) \comma \end{equation}
where 
$w = \pi (q-p)/(2N) $ ({\it c.f.} equation \ref{invsn} above and proposition 2 of
Ref.  \cite{KashMiwa}).

\subsection*{Chiral Potts model}

In 1987 Au-Yang, McCoy, Perk and others found solutions of the star-triangle
relations for the three- and four-state chiral Potts model \cite{AuYangetal87,
McCoyetal87}. These were generalized to an arbitrary number $N$ of states in
1988 \cite{BPAY88}.

In this model, the rapidity is a point $(a_p, b_p, c_p, d_p)$
on the homogeneous curve \cite{BPAY88}
\be
a_p^N+k' b_p^N \eq k \, d_p^N \sep k' a_p^N+ b_p^N \eq  k \, c_p^N \comma
\end{equation}
where $k$ and $k'$ are two given constants, related by
\be k^2 + {k'}^2 \eq 1 \period \end{equation}

Related quantities \cite{Baxter93} are $x_p, y_p, \mu _p , t_p$, defined by
 \be \label{xymu} 
x_p \eq a_p/d_p \sep y_p \eq b_p /c_p \sep \mu_p \eq d_p /c_p
\comma \end{equation}
\bd t_p \eq x_p y_p \eq a_p b_p /c_p d_p \period \ed 
They satisfy
\be \label{pvar}
x_q^N + y_q^N \eq  k(1 + x_q^N y_q^N) \sep k x_q^N \eq 1 - k'
\mu_q^{-N}  \sep k y_q^N \eq 1 - k' \mu_q^N   \period
\end{equation}

The chiral Potts model is $Z_N$-invariant, so we can write the Boltzmann weight
functions as in (\ref{zinvww}). Setting
$\omega = e^{2\pi i/N}$, the functions $W_{pq} (a - b) $, 
$\Wb _{pq} (a - b) $  are  
\be \label{WWbara} 
W_{pq} (n) \eq  \left( \frac{\mu_p }{ \mu_q } \right) ^n \prod_{j = 1}^{n} 
\frac{y_q - \omega^j x_p }{y_p - \omega^j x_q } \sep
\Wb _{pq} (n) \eq  (\mu_p  \mu_q )^n \prod_{j = 1}^{n} 
\frac{\omega x_p - \omega^j x_q }{y_q - \omega^j y_p } \period
\end{equation}
These Boltzmann weights are positive real if $x_p, x_q$, $y_p, y_q, \omega x_p$
all lie on the unit circle and are arranged sequentially in the widdershins
direction.

The function $f_{pq}$ has been  evaluated \cite{Baxter88, BBP90}. In
the present normalization, with $W_{pq} (0) = $ $\Wb_{pq} (0) = 1$, it is 
\be \label{fpqCP}
f_{pq}^N \eq  
\prod_{j=1}^{N-1} \left\{ \frac {\mu_q (1 - \omega^j ) (t_p-\omega^j t_q) (x_q -
\omega^j y_p )} { \mu_p (x_p - \omega^j x_q) (y_p - \omega^j y_q) (x_p -
\omega^j y_q) }
\right\}^j
\period \end{equation}

When $N = 2$ it reduces to the Ising model.

Let $R$ be the automorphism that takes the point $(a_p, b_p, c_p, d_p )$
to $(a_{Rp}$, $b_{Rp}, c_{Rp}, d_{Rp})$ $=$ $( b_p, \omega a_p, d_p, c_p )$.
Then the rotation relations (\ref{rotn}) are satisfied.  Using (\ref{defI}),
(\ref{frotn}) and (\ref{fpqCP}), we deduce that
\be \label{ICP}
I \eq f_{pq} \fb _{pq}/ N  \eq 1/{k'}^{(N-1)/N } \period \end{equation}
Thus $I$ is indeed rapidity-independent for the chiral Potts model,
being simply a power of $k'$. It is inverted by the duality relation
$k' \rightarrow 1/k'$ \cite[\S 5]{Bax89}.

The result (\ref{ICP}) is not new, being given in equation (2.47) of
Ref. \cite{BBP90}. What we have done here is show that its structure is a direct
consequence of the star-triangle relation, and is shared by other solvable models.

We can also deduce that
\be f_{pq} f_{qp} \eq \frac{N (t_p^N-t_q^N) (x_p - x_q) (y_p - y_q )}
{(x_p^N-x_q^N) (y_p^N-y_q^N) (t_p - t_q ) } \comma \end{equation}
in agreement with (2.48) of Ref. \cite{BBP90}.

It is not obvious from (\ref{fpqCP}) that the rhs is the  $N$th power of a 
single-valued function (to within factors that depend only on $p$, or only on $q$).
However, the only singularities whose location depends on both
 $p$ and $q$ occur at
the zeros and poles of the bracketted expressions. These can only occur
at certain points
$\cal P$ on the $(x_q, y_q)$ or $(x_p, y_p)$ Riemann surface, namely when
$x_q^N = x_p^N$ and $y_q^N = y_p^N$, or when $x_q^N = y_p^N$ and $y_q^N = x_p^N$.
We can count these zeros and poles by postulating the existence of functions
$\Theta_{ij}, \Thb_{ij}$ of $p$ and $q$  \cite{Bax93}, such that
$\Theta_{ij}$ has simple zeros when
\be
x_q \eq \omega^i y_p {\rm \; \; \; and \; \; \; } y_q \eq \omega^j x_p \comma \end{equation}
and $\Thb_{ij}$ has simple zeros when
\be
x_q \eq \omega^i x_p {\rm \; \; \; and \; \; \; } y_q \eq \omega^j y_p \period \end{equation}
Then, to within factors that are analytic and non-zero at all such points $\cal P$
\ba
x_q - \omega^i y_p \eq \prod_{m=0}^{N-1} \Theta_{i,m} & , & 
y_q - \omega^j x_p \eq \prod_{m=0}^{N-1} \Theta_{m,j} \nonumber \\
x_q - \omega^i x_p \eq \prod_{m=0}^{N-1} \Thb_{i,m} & , & 
y_q - \omega^j y_p \eq \prod_{m=0}^{N-1} \Thb_{m,j}  \ea
\bd
t_q - \omega^i t_p \eq  \prod_{m=0}^{N-1} \Theta_{m,i-m}
\Thb_{m,i-m} \period  \ed 
Formally substituting these expressions into (\ref{fpqCP}), we find that 
each $\Theta_{ij}$ or
$ \Thb_{ij}$ function occurs with a power $N, 0$ or $-N$ on the rhs, 
so we can take the 
$N$th root and obtain
\be f_{pq} \eq
\prod_{i=1}^{N-1} \, \prod_{j=1}^{N-i} \Theta_{N-i,i+j}\, /\, \Thb_{ij}
\comma
\end{equation}
 to within factors that are analytic and non-zero at all  points $\cal P$.
So $f_{pq}$ certainly has no branch point singularities at the points $\cal
P$: at worst it has simple poles or zeros.

\subsection*{Fateev-Zamolodchikov model}

When $k = 0$ we can choose the parameters of the chiral Potts model so that
\be
x_p \eq e^{i \pi p /N} \sep y_p \eq e^{i \pi (p+1) /N} \sep \mu_p \eq 1 
\comma 
\end{equation}  
where $p$ is the rapidity variable. Making the corresponding choices for $x_q, y_q,
\mu_q$, (\ref{WWbara}) becomes 
\ba \label{WFZ}
W_{pq} (n) & = & \prod_{j=1}^n \, \frac{\sin [\pi (p - q + 2 j-1)/2 N ] }{
\sin [\pi (q - p + 2 j-1)/2 N ]} \comma \\ 
\Wb_{pq} (n) & = & \prod_{j=1}^n \, \frac{\sin [\pi (q-p + 2 j-2)/2 N ] }{
\sin [\pi (p - q + 2 j)/2 N ]} \period \nonumber \ea
We obtain precisely the same result if we set $k=0$ (and hence $\tilde{q} = 0$) in
the Kashiwara-Miwa model, so the two models then coincide.

The resulting model is the Fateev-Zamolodchikov model \cite{FatZam82}. Its weights
(\ref{WFZ}) are real and positive if $0 < q-p < 1$. It is a $Z_N$ model, self-dual and
critical. When $N=3$ it is equivalent to the self-dual Potts model (with $\mu =
\pi/6$ and $p, q$ scaled by $\mu$).

From both (\ref{IKM}) and (\ref{ICP}) we deduce that
\be I \eq 1 \comma \end{equation}
in agreement with (\ref{selfdual}).

We find that
the right-hand side of the chiral Potts result (\ref{fpqCP}) is now indeed a perfect
$N$th power, and that
\be \label{FZfpq}
f_{pq} \eq N^{1/2} \, \prod_{j=1}^{N'} \frac{\sin [\pi (q-p+2 j-1)/2 N ] }
{\sin [ \pi (p - q + 2 j)/2 N] } \comma \end{equation}
where $N' = [N/2]$. This agrees with, and is part of the evidence for, the
conjecture (\ref{KMfpq}) for the Kashiwara-Miwa model. 

\section{Summary}

The method of Matveev and Smirnov \cite{MatSmirn90} can be used to show that the quantity $I$
defined by (\ref{deffpq}), (\ref{deffbpq}), (\ref{defI}) is independent of the rapidities $p$
and $q$. For particular models this is not necessarily a new observation, but the realisation
that it is a direct consequence of the star-triangle relation does provide a unifying feature
for  the known solvable edge-interaction models. We hope this is a small step forward in the
subject to which Barry McCoy has made so many outstanding contributions.


\bibliographystyle{plain}

\vskip 2pc
{\noindent \em R.J. Baxter,
 Department of Theoretical Physics and School of Mathematical Sciences,
I.A.S., The Australian National University,
Canberra  A.C.T. 0200, Australia,
 e-mail: {\tt rj.baxter@anu.edu.au }
    }


}

\end{document}